\documentclass[10pt]{article}
\pdfoutput=1 

\usepackage[preprint]{tmlr}
\usepackage{rawfonts, amsmath, amsfonts, graphicx, caption, subfig, makecell, lastpage, algorithm, algpseudocode, upgreek}
\usepackage[hidelinks]{hyperref}
\usepackage{url}

\usepackage{fancyhdr}
\DeclareUnicodeCharacter{2212}{-}

\begin{document}
\title{Multidimensional Gabor-Like Filters Derived from Gaussian Functions on Logarithmic Frequency Axes}

\author{\name Dherik Devakumar \email dherikjd@stud.ntnu.no\\ \addr Norwegian University of Science and Technology,\\ Department of Computer Science,\\ NO-7491 Trondheim, Norway \AND Ole Christian Eidheim \email ole.c.eidheim@ntnu.no\\ \addr Norwegian University of Science and Technology,\\ Department of Computer Science,\\ NO-7491 Trondheim, Norway}

\maketitle


\begin{abstract}
A novel wavelet-like function is presented that makes it convenient to create filter banks given mainly two parameters that influence the focus area and the filter count. This is accomplished by computing the inverse Fourier transform of Gaussian functions on logarithmic frequency axes in the frequency domain. The resulting filters\footnote{The source code to reproduce the figures in this article can be found at https://gitlab.com/eidheim/gabor-like-filters.} are similar to Gabor filters and represent oriented brief signal oscillations of different sizes. The wavelet-like function can be thought of as a generalized Log-Gabor filter that is multidimensional, always uses Gaussian functions on logarithmic frequency axes, and innately includes low-pass filters from Gaussian functions located at the frequency domain origin.
\end{abstract}


\section{Introduction}


Most previous wavelets and wavelet-like functions do not fully take into account the exponential nature of the frequency domain, which is evident in musical octaves. For example, the popular Gabor filters \citep{gabor1946theory} are isotropic Gaussian functions in the frequency domain, and when these functions are shifted along a frequency axis, the resulting filters are not only scaled, but changed in a non-affine manner. To reduce these transformations, filters can instead be derived from Gaussian functions on logarithmic frequency axes in the frequency domain. Lower frequencies will then be covered more narrowly, while higher frequencies will be more widely covered.

Logarithmically scaled frequencies are used in the Log-Gabor filter \citep{field1987edge} and its extension to 2D by \citet{fischer2007self}. However, the original Log-Gabor filter is not defined for negative frequencies, and the 2D extension does not result in Gaussian functions on logarithmic frequency axes. Furthermore, the presented wavelet-like function has fewer parameters than the Log-Gabor 2D extension, and is defined for any number of dimensions.

The function is called wavelet-like since it does not satisfy all of the wavelet criteria given in \citet{valens1999really,sheng2010wavelet}, in particular the admissibility condition due to non-zero mean of filters derived from Gaussian functions close to the frequency domain origin. On the other hand, these criteria may not be decisive with respect to the application of a filter bank, and in fact the filter from a Gaussian function at the frequency domain origin corresponds to a low-pass filter commonly used in conjunction with Gabor filters and other wavelet or wavelet-like functions.

\section{Continuous Function}

Let $\boldsymbol{x},\boldsymbol{\xi} \in \mathbb{R}^D$ be spatial coordinates and frequencies, respectively. The wavelet-like function is then defined as the inverse Fourier transform of a Gaussian function on logarithmic frequency axes, given center $\boldsymbol{\mu} \in \mathbb{R}^D$ and width $\sigma \in \mathbb{R}_{>0}$:

\[
  \psi_{\boldsymbol{\mu},\sigma}(\boldsymbol{x}) =
    \int_{\mathbb{R}^D}
      e^{-{\left \lVert \boldsymbol{\xi}/{\lVert\boldsymbol{\xi}\rVert}_2 \ln({\lVert\boldsymbol{\xi}\rVert}_2 + 1)- \boldsymbol{\mu} \right \rVert}_{2}^2 / \sigma}
      e^{2\pi i \boldsymbol{x} \cdot \boldsymbol{\xi}} \textrm{d}\boldsymbol{\xi}
\]

where ${\lVert \ \rVert}_{2}$ denotes the $l_{2}$-norm and $\cdot$ represents the dot product. The frequencies $\boldsymbol{\xi}$ are factored into vector direction $\frac{\boldsymbol{\xi}}{{\lVert \boldsymbol{\xi} \rVert}_2}$ and length ${\lVert \boldsymbol{\xi} \rVert}_2$, of which the length is scaled logarithmically into $[0, \infty)$. The addition of $1$ in the $ln$-expression ensures Gaussian functions, although compressed or stretched on regular axes, for all $\boldsymbol{\mu}$. In the case of $\boldsymbol{\xi} = \boldsymbol{0}$, $\frac{\boldsymbol{\xi}}{{\lVert \boldsymbol{\xi} \rVert}_2} \ln({\lVert\boldsymbol{\xi}\rVert}_2 + 1)$ is defined to be $\lim_{\boldsymbol{\xi} \to \boldsymbol{0}} \frac{\boldsymbol{\xi}}{{\lVert \boldsymbol{\xi} \rVert}_2} \ln({\lVert\boldsymbol{\xi}\rVert}_2 + 1) = \boldsymbol{0}$.

Considering the real and imaginary parts separately and $\operatorname{Im}(\psi_{\boldsymbol{\mu},\sigma}(\boldsymbol{x})) = 0 \ \forall \boldsymbol{x}$ if $\boldsymbol{\mu} = \boldsymbol{0}$, normalization of $\psi$ can be defined as:

\[
  \tilde{\psi}_{\boldsymbol{\mu},\sigma}(\boldsymbol{x})=
  \begin{cases}
    \frac{\operatorname{Re}(\psi_{\boldsymbol{\mu},\sigma}(\boldsymbol{x}))}{\sqrt{\int_{\mathbb{R}^D} {\operatorname{Re}(\psi_{\boldsymbol{\mu},\sigma}(\boldsymbol{x}))}^2 \textrm{d}\boldsymbol{x}}},& \text{if } \boldsymbol{\mu} = \boldsymbol{0}\\
    \frac{\operatorname{Re}(\psi_{\boldsymbol{\mu},\sigma}(\boldsymbol{x}))}{\sqrt{\int_{\mathbb{R}^D} {\operatorname{Re}(\psi_{\boldsymbol{\mu},\sigma}(\boldsymbol{x}))}^2 \textrm{d}\boldsymbol{x}}} + \frac{\operatorname{Im}(\psi_{\boldsymbol{\mu},\sigma}(\boldsymbol{x}))}{\sqrt{\int_{\mathbb{R}^D} {\operatorname{Im}(\psi_{\boldsymbol{\mu},\sigma}(\boldsymbol{x}))}^2 \textrm{d}\boldsymbol{x}}} i,& \text{otherwise}
  \end{cases}
\]

and $\tilde{\psi}$ will then have the following properties:

\[
  \begin{aligned}
    \int_{\mathbb{R}^D} {\lvert \tilde{\psi}_{\boldsymbol{\mu},\sigma}(\boldsymbol{x}) \rvert}^2 \textrm{d}\boldsymbol{x} &=
      \begin{cases}
        1,& \text{if } \boldsymbol{\mu} = \boldsymbol{0}\\
        2,& \text{otherwise}
      \end{cases}
      \\
    \int_{\mathbb{R}^D} {\operatorname{Re}(\tilde{\psi}_{\boldsymbol{\mu},\sigma}(\boldsymbol{x}))}^2 \textrm{d}\boldsymbol{x} &= 1\\
    \int_{\mathbb{R}^D} {\operatorname{Im}(\tilde{\psi}_{\boldsymbol{\mu},\sigma}(\boldsymbol{x}))}^2 \textrm{d}\boldsymbol{x} &=
      \begin{cases}
        0,& \text{if } \boldsymbol{\mu} = \boldsymbol{0}\\
        1,& \text{otherwise}
      \end{cases}
  \end{aligned}
\]

Due to the orthogonality of $\operatorname{Re}(\psi)$ and $\operatorname{Im}(\psi)$, the energy of $\tilde{\psi}$ is constant unless $\boldsymbol{\mu}$ is positioned at the frequency domain origin. The imaginary parts of the results from $\boldsymbol{\mu} = \boldsymbol{0}$ are all zero and can be discounted.

\section{Discrete Function}

By setting $N$ as an odd positive number and $\boldsymbol{n},\boldsymbol{k} \in \{-\frac{N-1}{2}, \ldots, \frac{N-1}{2}\}^D$, the wavelet-like function can be written discretely as:

\[
  \Psi_{\boldsymbol{\mu},\sigma}(\boldsymbol{n}) =
    \sum_{\boldsymbol{k} \in \{-\frac{N-1}{2}, \ldots, \frac{N-1}{2}\}^D}
      e^{-{\left \lVert s(N) \boldsymbol{k}/{\lVert\boldsymbol{k}\rVert}_2 \ln({\lVert\boldsymbol{k}\rVert}_2 + 1)- \boldsymbol{\mu} \right \rVert}_{2}^2 / \sigma}
      e^{2\pi i \boldsymbol{n} \cdot \boldsymbol{k}/N}
\]

where $s(N)=\frac{N-1}{2 ln(\frac{N+1}{2})}$ is used to normalize the logarithmic frequency axes such that their maxima are equal to that of the regular frequency axes. In two dimensions, i.e. $D=2$, $\Psi$ represents an $N \times N$ matrix of complex numbers. These numbers can constitute two filters in the context of image processing, one filter from the real parts and the second filter from the imaginary parts. As in $\psi$, $\frac{\boldsymbol{k}}{{\lVert \boldsymbol{k} \rVert}_2} \ln({\lVert\boldsymbol{k}\rVert}_2 + 1)$ is defined to be $\boldsymbol{0}$ if $\boldsymbol{k} = \boldsymbol{0}$.

In order to avoid sharp amplitude cutoffs that cause ripple effects in the spatial domain, the inherent symmetry of the frequency domain is exploited by adding the surplus magnitudes as follows:

\begin{equation}
  \label{equation1}
  \Uppsi_{\boldsymbol{\mu},\sigma}(\boldsymbol{n}) =
    \sum_{\boldsymbol{k} \in \{-\frac{N-1}{2}, \ldots, \frac{N-1}{2}\}^D}
      \left ( \sum_{\boldsymbol{l} \in \{-N, 0, N\}^D}
        e^{-{\left \lVert s(N) (\boldsymbol{k} + \boldsymbol{l})/{\lVert \boldsymbol{k} + \boldsymbol{l} \rVert}_2 \ln({\lVert \boldsymbol{k} + \boldsymbol{l} \rVert}_2 + 1)- \boldsymbol{\mu} \right \rVert}_{2}^2 / \sigma} \right )
      e^{2\pi i \boldsymbol{n} \cdot \boldsymbol{k}/N}
\end{equation}

\begin{figure}[tb]
  \setlength{\tabcolsep}{1.5pt}
  \centering
  \begin{tabular}{llll}
    \subfloat[][\label{1a}{\begin{tabular}[t]{c}Frequency domain\end{tabular}}]{\includegraphics{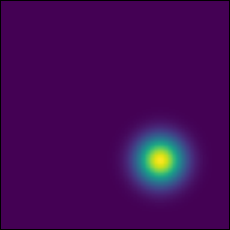}} &
    \subfloat[][\label{1b}{\begin{tabular}[t]{c}Frequency domain\end{tabular}}]{\includegraphics{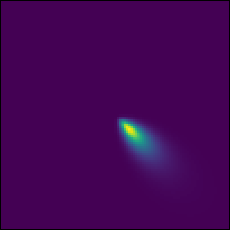}} &
    \subfloat[][\label{1c}{\begin{tabular}[t]{c}Spatial domain,\\real parts\end{tabular}}]{\includegraphics{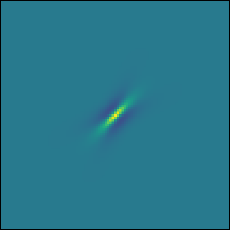}} &
    \subfloat[][\label{1d}{\begin{tabular}[t]{c}Spatial domain,\\imaginary parts\end{tabular}}]{\includegraphics{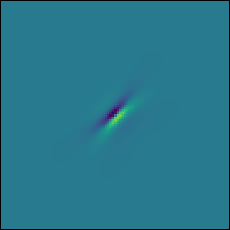}}
  \end{tabular}
  \caption{A demonstration of Equation \ref{equation1} with centered domain origins. The Gaussian function shown in a) is placed on logarithmic frequency axes as can be seen in b) using regular axes. The results of $\Uppsi$ shown in c) and d) are approximately orthogonal and sum to roughly zero.}
  \label{figure1}
\end{figure}

\citet{fischer2007self} further describe a half-pixel shift operation to improve the high-frequency imaginary results, however, these results would no longer be centered at the spatial domain origin. Therefore, $\Uppsi$ results from $\boldsymbol{\mu}$ elements close to $\pm \frac{N-1}{2}$, which represent the highest frequency fluctuations, are instead disregarded. A demonstration\footnote{An interactive demonstration is available at: \url{https://gitlab.com/eidheim/gabor-like-filters/-/blob/main/demo.py}.} of Equation \ref{equation1} is shown in Figure \ref{figure1} where $D=2$, $N=101$, $\boldsymbol{\mu}=(20, 20)$, and $\sigma=100$.

\section{Filter Bank}\label{filter_bank}

\begin{figure}[t]
  \setlength{\tabcolsep}{1.5pt}
  \centering
  \begin{tabular}{llllllll}
    \includegraphics{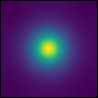} & \includegraphics{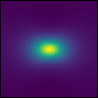} & \includegraphics{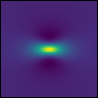} & \includegraphics{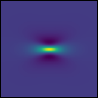} & \includegraphics{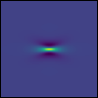} & \includegraphics{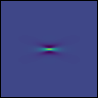} & \includegraphics{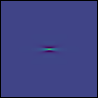} & \includegraphics{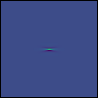} \\
    & \includegraphics{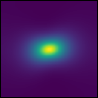} & \includegraphics{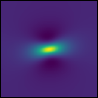} & \includegraphics{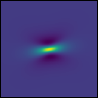} & \includegraphics{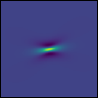} & \includegraphics{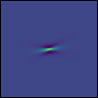} & \includegraphics{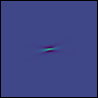} & \includegraphics{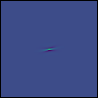} \\
    & \includegraphics{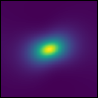} & \includegraphics{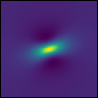} & \includegraphics{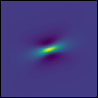} & \includegraphics{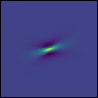} & \includegraphics{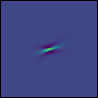} & \includegraphics{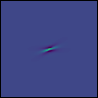} & \includegraphics{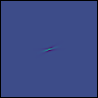} \\
    & \includegraphics{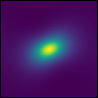} & \includegraphics{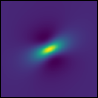} & \includegraphics{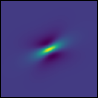} & \includegraphics{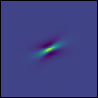} & \includegraphics{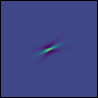} & \includegraphics{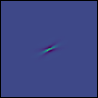} & \includegraphics{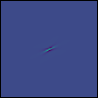} \\
    & \includegraphics{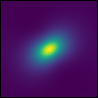} & \includegraphics{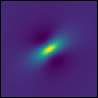} & \includegraphics{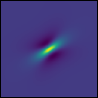} & \includegraphics{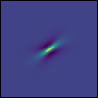} & \includegraphics{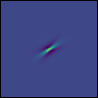} & \includegraphics{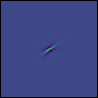} & \includegraphics{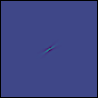} \\
    & \includegraphics{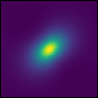} & \includegraphics{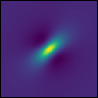} & \includegraphics{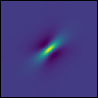} & \includegraphics{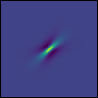} & \includegraphics{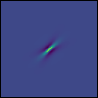} & \includegraphics{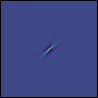} & \includegraphics{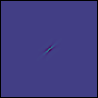} \\
    & \includegraphics{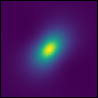} & \includegraphics{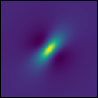} & \includegraphics{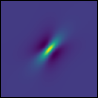} & \includegraphics{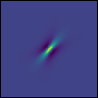} & \includegraphics{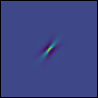} & \includegraphics{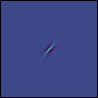} & \includegraphics{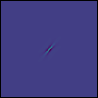} \\
    & \includegraphics{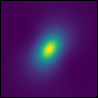} & \includegraphics{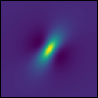} & \includegraphics{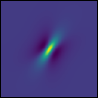} & \includegraphics{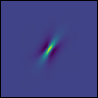} & \includegraphics{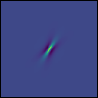} & \includegraphics{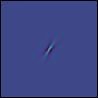} & \includegraphics{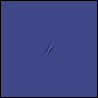} \\
    & \includegraphics{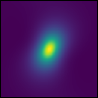} & \includegraphics{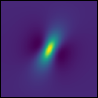} & \includegraphics{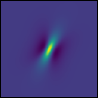} & \includegraphics{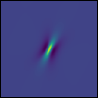} & \includegraphics{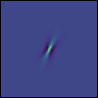} & \includegraphics{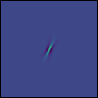} & \includegraphics{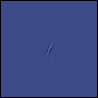} \\
    & \includegraphics{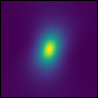} & \includegraphics{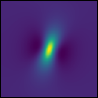} & \includegraphics{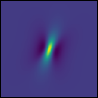} & \includegraphics{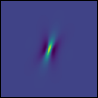} & \includegraphics{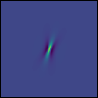} & \includegraphics{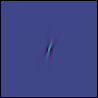} & \includegraphics{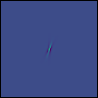} \\
    & \includegraphics{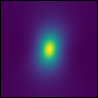} & \includegraphics{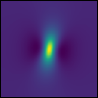} & \includegraphics{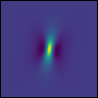} & \includegraphics{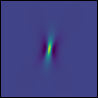} & \includegraphics{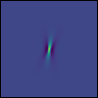} & \includegraphics{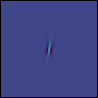} & \includegraphics{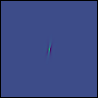} \\
    \includegraphics{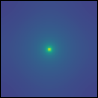} & \includegraphics{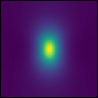} & \includegraphics{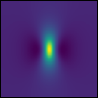} & \includegraphics{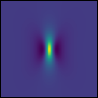} & \includegraphics{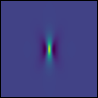} & \includegraphics{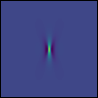} & \includegraphics{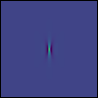} & \includegraphics{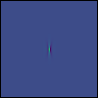} \\
  \end{tabular}
  \caption{The real parts of an example filter bank from Equation \ref{equation1}, where the top left image shows a low-pass filter from $\boldsymbol{\mu} = \boldsymbol{0}$. The bottom left image displays the sum of the Gaussian functions in the frequency domain over $\boldsymbol{\mu} \in \{0, 0\} \cup \{(r \cos \theta, r \sin \theta)\ |\ r \in \{6, 12, \ldots, 48\}$, $\theta \in \{0, \frac{\pi}{22}, \frac{2\pi}{22}, \ldots, \frac{43\pi}{22}\}\}$, which shows the potential frequency domain coverage of the utilized $r$ and $\theta$ intervals. Moreover, the normalized real parts of the inverse Fourier transform of the bottom left image approximate an identity filter, i.e. $1$ at the origin and $0$ elsewhere.}
  \label{figure2}
\end{figure}

The real parts of an example filter bank from Equation \ref{equation1} are shown in Figure \ref{figure2} with $D=2$, $N=101$, and $\sigma=100$. The Gaussian centers are set regularly along the first axis and rotated around the origin of the frequency domain:

\[
  \boldsymbol{\mu} \in \{(0,0)\} \cup \{(r \cos \theta, r \sin \theta)\ |\ r \in \{6, 12, \ldots, 42\}, \theta \in \{0, \frac{\pi}{22}, \frac{2\pi}{22}, \ldots, \frac{\pi}{2}\}\}
\]

The change in $\theta$ is set such that the distance between the outer rotated centers is approximately equal to the change in $r$ along the first axis:

\[
  {\lVert (42 \cos \theta, 42 \sin \theta) - (42, 0)\rVert}_2 = 6
\]

of which one solution is $\theta \approx \frac{\pi}{22}$.

\bibliography{article}
\bibliographystyle{tmlr}
\end{document}